\documentclass{article}
\usepackage[T1]{fontenc}
\usepackage{geometry}
\usepackage{graphicx}
\usepackage{authblk}
\usepackage{xcolor}
\usepackage{url}
\usepackage{algpseudocode}  
\usepackage{algorithm}
\usepackage{amsmath}        
\usepackage{amsfonts}       
\usepackage{amsthm}
\usepackage{hyperref} 

\usepackage[a-2u]{pdfx}
\usepackage{subcaption}
\usepackage{placeins}

\theoremstyle{plain}

\begin{document}
\title{Local iterative algorithms for approximate symmetry guided by network centralities}
%
%
\author[1]{David Hartman}
\author[1]{Jaroslav Hlinka}
\author[1]{Anna Pidnebesna}
\author[1]{Franti\v{s}ek Szczepanik}

\affil[1]{Institute of Computer Science of the Czech Academy of Sciences, \\ Pod Vod\'{a}renskou v\v{e}\v{z}\'{i} 271/2, 182 07 Prague, Czech Republic}
\maketitle              
\begin{abstract}
Recently, the influence of potentially present symmetries has begun to be studied in complex networks. A typical way of studying symmetries is via the automorphism group of the corresponding graph. Since complex networks are often subject to uncertainty and automorphisms are very sensitive to small changes, this characterization needs to be modified to an approximate version for successful application. This paper considers a recently introduced approximate symmetry of complex networks computed as an automorphism with acceptance of small edge preservation error, see Liu 2020~\cite{liu2020approximate}. This problem is generally very hard with respect to the large space of candidate permutations, and hence the corresponding computation methods typically lead to the utilization of local algorithms such as the simulated annealing used in the original work. This paper proposes a new heuristic algorithm extending such iterative search algorithm method by using network centralities as heuristics. Centralities are shown to be a good tool to navigate the local search towards more appropriate permutations and lead to better search results.

\end{abstract}
\section{Introduction}
\addcontentsline{toc}{chapter}{Introduction}

Complex networks are graphs that represent various natural and man-made systems such as, for example, the Internet~\cite{internet-betweenness}, public transportation~\cite{transportation}, or the human brain~\cite{brain}. A typical way of analyzing such systems is to use several characteristics, both local (bound to vertices or edges) and global (bound to the whole graph). One notable global property is network symmetry, which received attention since MacArthur et al.~\cite{MacArthur} showed that many real-world networks contain high symmetry. Network symmetry has proven useful in revealing patterns of synchronized node clusters, helping understand various forms of collective behavior~\cite{synchronized-node-clusters}.

All types of network characteristics can be burdened with uncertainty problems. This uncertainty can be due to the construction of the corresponding network, an example is the uncertainty of local characteristics caused by nonlinearity in time series, see~\cite{Hartman2018nonlinearity,Hlinka2014nonlinear,Hartman2011role}. The second type of uncertainty is the sensitivity of the corresponding characteristics to the overall behavior of the system, exemplified by the sensitivity of the global characteristic called a small-world coefficient, see~\cite{hlinka2012small,Hlinka2017small,small-world}. Both types of uncertainty are combined in the graph symmetry represented by the automorphic group mainly because of its sensitivity to small changes.

For this reason, a more robust approximate symmetry based on the distance to the originally introduced automorphic symmetry was recently introduced, see Liu 2020~\cite{liu2020approximate}. The proposed numerical solution was based on simulated annealing omitting the fixed points of the corresponding permutation. In a recent improvement of this method, see~\cite{pidnebesna2023computing}, this approach has been extended by introducing fixed points and also by using a relaxed problem and a subsequent continuous optimization.


In practice, simulated annealing struggles to find an optimal permutation in the factorially large solution space. To address these limitations, we introduce a new heuristic method. This heuristic is based on the observation that permutations close to automorphisms tend to display similar vertices to themselves. We measure vertex similarity using graph centralities such as eigenvector centrality, PageRank, or betweenness. We compare the performance of the original and heuristic algorithm on several random network models, building on the experiments carried out by Pidnebesna et al.~\cite{pidnebesna2023computing}. Presented results are based on the results of the bachelor thesis of F. Szczepanik, where one can find more detailed descriptions of the basics of the methods used and some additional observations~\cite{szczepanik2024centralities}.

\section{Preliminaries}
\subsection{Basic notation and definitions.}
The object of interest is a graph (undirected, unweighted, loop-less), defined by a set of vertices (nodes) $V$ and edges (links) $E$, denoted as $G = (V, E)$. We use the standard notion of an adjacency matrix denoted as $A$, where $a_{ij} = 1$ if $\{v_i, v_j\} \in E$ and $a_{ij} = 0$ otherwise. We denote the set of vertices adjacent to a given vertex as $N_i = \{v_j | \{v_i, v_j\} \in E\}$. We call its size as $k_i = |N_i|$ and call it the degree of a node $i$. For a set of nodes $U \subseteq V$ we denote $E(U)$ as the set of edges induced by $U$, i.e. set of edges where each edge has at least one node in $U$. The Cartesian product $G \times H$ of two graphs $G$ and $H$ is a graph such that 1) $V(G \times H) = V(G) \times V(H)$, and 2) $\{(u, u'), (v, v')\} \in E(G \times H)$ if $u = v$ and $\{u',v'\} \in E(H)$ or $u' = v'$ and $\{u,v\} \in E(G)$.

An automorphism of a graph $G = (V, E)$ is a permutation $\pi$ of $V$ preserving adjacencies, i.e.: $\{\pi(v_x),\pi(v_y)\} \in E \iff \{v_x, v_y\} \in E$.
We represent such permutations via matrices: the permutation matrix $P$ for a permutation $\pi$ on $n$ elements is an $n \times n$ matrix where:
\[
P_{ij} = 
\begin{cases} 
1 & \text{if } i = \pi(j), \\
0 & \text{otherwise.}
\end{cases}
\]
$PA$ permutes $A$'s rows and $AP$ permutes its columns according to $\pi$. Considering the orthogonality of permutation matrices, we obtain the following~\cite{Biggs_1974}: {\em the permutation $\pi$ is an automorphism of $G$ with an adjacency matrix $A$ if and only if $A = PAP^{\mathrm{T}}$}. Lastly, the L1 norm of an $n \times m$ real matrix $A$ is: $
 \|A\|_{1}={{\sum _{i}^{m}\sum _{j}^{n}|a_{ij}|^{2}}}$.

\subsection{Approximate Symmetry}


The approximate symmetry of a graph $G = (V,E)$ with an adjacency matrix $A$ is defined as~\cite{liu2020approximate}:

\[
E(A) = \frac{1}{4} \min_{P} (\left\| A - PAP^{\mathrm{T}} \right\|_{1}).
\]

$E(A)$ minimizes the number of mismatches $A - PAP^{\mathrm{T}}$ over permutation matrices $P$. Due to the constant $\frac{1}{4}$, it directly expresses the number of mismatched edges that differ after the permutation: $A_{ij} \neq A_{\pi(i)\pi(j)}$.
So, if $A = PAP^{\mathrm{T}}$, then $P$ corresponds to an automorphism, and $E(A)$ is equal to zero.

A parametrized symmetry version $\epsilon(A,P)$ is then defined as follows:

\[
\epsilon(A, P) = \frac{1}{4}\left\| A - PAP^{\mathrm{T}} \right\|_{1}
,\]
yielding $E(A) = \min_{P} \epsilon(A, P)$. For comparability across graphs, Liu introduced \emph{normalized symmetry} $S(A)$, which normalizes $E(A)$ by the maximal $\epsilon(A, P)$. The maximal value of $\epsilon(A,P)$ is reached when the graph has exactly half of all possible edges, that is $\frac{1}{2} {n \choose 2}$, and all of them are mismatched by $P$. The normalized approximate symmetry of a graph $G=(V,E)$ with an adjacency matrix $A$ is thus:

\[
S(A) = \frac{E(A)}{\frac{1}{2}{n \choose 2}} = \frac{\min_{P} \left(\left\| A - PAP^{\mathrm{T}} \right\|_{1}\right)}{n(n-1)}
.\]

Identifying the optimal approximate automorphism $P$ minimizing $\epsilon(A, P)$ is computationally hard as it involves searching through a factorially large solution space of vertex permutations. One of the suitable metaheuristics for this setting is Simulated Annealing, a probabilistic optimization technique~\cite{simanneal}. Annealing explores the solution space by transitioning from one state to another while decreasing \emph{energy}, which represents the value of the objective function. The process is initialized at a random state and initially explores worse states to avoid getting stuck in local optima according to the current \emph{temperature}. The temperature is gradually reduced according to a chosen \emph{cooling schedule} until some threshold temperature is reached or all move iterations are carried out.

\subsection{Graph Centralities}



Graph centralities are characteristics that evaluate the importance of a vertex from a structural perspective. More details about centralities can be found e.g. in~\cite{Newman}. One of the simplest measures is {\em degree centrality} defined as the number of adjacent edges. This centrality captures mainly local effects around the vertices. The generalization of the node degree is {\em eigenvector centrality}, which considers the importance of the vertex neighbors in addition to the number of its neighbors. Moreover, we apply this consideration to more distant vertices. This centrality can be defined recursively. If we have defined the initial centrality as $e_i(0)$ defined for a node $i$, then the total eigenvector centrality for this node can be defined as 
\begin{equation*}
e_i(t) = \sum_{j=1}^n A_{ij} e_j(t-1).
\end{equation*}
The exact value of the centrality is computed for $t \rightarrow \infty$.  It is possible to show further that this equation can be rewritten into the following characteristic equation $e_i = \frac{1}{\lambda} \sum_{j} A_{ij} k_j$, where $\lambda$ is the greatest eigenvalue of $A$. Thus eigenvector centrality can be expressed as the eigenvector corresponding to the largest eigenvalue of $A$~\cite{Newman}. The last generalization in this direction is {\em PageRank}, which originated as a system for ranking the importance of web pages~\cite{PageRank}. The main generalization of this centrality over the eigenvector centrality is the normalization of steps with degrees 
\begin{equation*}
e_i(t) = \alpha\sum_{j=1}^n A_{ij} \frac{e_j(t-1)}{k_j},
\end{equation*}
where $\alpha$ is a damping
factor typically set to 0.85, see full conditions in~\cite{Newman}.



Instead of studying the number of neighbors, one can also study the density of edges in the neighborhood. This is measured by the {\em clustering coefficient} $c_i$ of a node $i$ defined as 
\begin{equation*}    
c_i = \frac{2 |E(N_i)|}{|N_i| |N_i - 1|}.
\end{equation*}

This centrality measures how locally clustered the neighborhood of a vertex is, i.e. how much the vertex is part of the denser graphs~\cite{NetworkScience}.

The last centrality is representative of a class of characteristics measuring the role of a vertex in communication through the network. In more detail, {\em betweenness centrality} assesses a node's importance by counting the number of shortest paths between all pairs of nodes passing through it~\cite{pokorna}.
The betweenness centrality of a node $v_i$ is defined as:
\[
b_i = \sum_{\substack{x,y \in V(G) \\ x \neq y \neq v_i}} \frac{\sigma_{xy}(v_i)}{\sigma_{xy}},
\]
where $\sigma_{xy}$ denotes the number of shortest paths between nodes $x$ and $y$ and $\sigma_{xy}(v_i)$ is the number of shortest paths between $x$ and $y$ passing through $v_i$~\cite{pokorna}.


\subsection{Random network models}
Random network models provide abstractions of real-world systems and aim to capture their structural properties. 
In this section, we introduce several random graphs and random network models that serve as the dataset for comparing and evaluating the original and heuristic versions of simulated annealing. For such testing, we follow the methodologies of Straka~\cite{Straka} and Pidnebesna et al.~\cite{pidnebesna2023computing}.

The first class of graphs is not random but suitable for symmetry testing due to its high regularity. 
A {\em grid graph} in $d$ dimensions and lengths $n_i$ in dimension $i$ is defined as a graph product of path graphs $P_{n_1} \times P_{n_2} \times ... \times P_{n_d}$ and denoted as $R_{n_1, n_2,...,n_d}$~\cite{pidnebesna2023computing}. 

The simplest random graph is the standard {\em Erd\H{o}s–Rényi} model (ER)~\cite{er-model}. An ER graph $G(n,p)$ is generated on $n$ vertices by adding an edge between every pair with probability $p$. The ER graph is simple to analyze but does not represent the real networks very accurately, e.g. lacks the small-world as well as scale-free character~\cite{small-world,Newman}. 

A more realistic model is the {\em Barabási-Albert} model (BA) that represents network growing models~\cite{NetworkScience}.  The generation procedure for the selected parameters $n$ and $k$ is as follows. We start with a small ($m \ll n$) graph and then add a vertex at each step, which we connect to $k$ existing vertices, preferring higher degree vertices, namely the probability of connecting to an existing node $i$ is $p_i = k_i / \sum_j k_j$. This model captures some characteristics such as power-law degree distribution (scale-free) and the emergence of important high-degree nodes (hubs)~\cite{NetworkScience}.

The {\em duplication-divergence} (DD) model simulates the evolution of protein-protein interaction (PPI) networks. PPI networks evolve by the mechanism of gene duplication, which generates new proteins that are initially identical to pre-existing ones but gradually diverge, with most of the new proteins not surviving. The duplication–divergence is described by the following growth mechanism~\cite{DDNetworks}:
\begin{itemize}
    \item Duplication: A node is randomly selected and duplicated. The newly created node inherits all links of the duplicated node.
    \item Divergence: Every link of the new node is retained with \emph{divergence probability} $\sigma$. If no links survive, the duplication is unsuccessful, and the new node is discarded.
\end{itemize}

This model represents a real network but has some form of symmetry directly embedded in the generative process.



\section{Heuristic method}

Our primary objective is to investigate how guiding annealing with graph centralities improves the detection of approximate network symmetry. 
Thus, in this study, we introduce a modified annealing algorithm, \emph{guided annealing}, which uses centrality metrics as a heuristic for selecting transpositions when moving across states. 

Roughly speaking, a basic approach to exploring the solution space of approximate automorphisms involves applying random transpositions to the current permutation. Enhancing the transition strategy by incorporating some form of guidance could improve the efficiency of the annealing process.
An optimal approximate automorphism should align vertices with similar structural roles in the graph. For instance, mapping a hub node onto a peripheral node will likely result in a large number of mismatched edges and therefore suboptimal approximate symmetry. Graph centralities allow us to quantify vertex “similarity” and thus identify vertices that should be aligned.
An important property of graph centralities is that \emph{automorphisms preserve centralities} ~\cite{pokorna,pagerankautomorphism}. Therefore, we expect vertices with similar centrality values to be aligned in an optimal permutation.

\subsection{Move function in guided annealing}
We reimplement the \emph{move} function, which generates a new permutation $\pi'$ from the current permutation $\pi$ and thus determines the next state. In unmodified simulated annealing, the transposition is constructed by randomly swapping the images of two vertices under the current permutation (moving from $a,\pi(a)$ and $b, \pi(b)$ to $a,\pi(b)$ and $b,\pi(a)$, where $a$ and $b$ are vertices and $\pi$ the current permutation). In our guided version, the transpositions aligning similar vertices are chosen with higher probability.

\subsubsection{Similarity matrix and implementation description}

We express the similarity of vertex pairs for a given centrality $\Gamma$ in an $n \times n$ matrix $M$, indexed by vertices.
To construct $M$, we start by computing a \emph{difference matrix} $D$, where each element is defined as $d_{ij} = |\Gamma(i) - \Gamma(j)|$, i.e. the absolute difference of the centrality of $v_i$ and $v_j$. To transform this into a similarity measure where higher values imply higher similarity, we compute inverses over the elements of $D$. Specifically, the entries of $M$ are defined as $m_{ij} = (d_{ij} + \beta)^{-1}$, where $\beta > 0$ is a division constant preventing the denominator from being zero and moderating the variance of the matrix values; as $\beta$ increases, the entries of $M$ become more uniform, guiding the annealing less aggressively.

\subsubsection{Re-implementing the move function}
We now describe the re-implemented move function. We start by randomly choosing the first vertex $a$ \footnote{We also experimented with a “one-step” approach, which does not select the first vertex randomly but instead considers all possible transpositions. However, this requires $O (n^{2})$ evaluations in each step or a one-time pre-computation for all pairs of pairs, which has a complexity of $O(n^{4})$, both practically unusable for growing graphs.}, where $\pi(a)$ is its image under the current permutation and $m_{a\pi(a)}$ is their similarity. We evaluate a potential similarity increase by considering swaps with each vertex $b$ and its image $\pi(b)$. The difference in the similarity of the original and proposed setting can be calculated as $\Delta M_b = m_{a\pi(b)} + m_{b\pi(a)} - m_{a\pi(a)} - m_{b\pi(b)}$.

Before normalizing $\Delta M$ into a probability distribution over vertices, we again introduce a smoothing parameter $\phi > 0$, which we call the probability constant, and set $\Delta E_b = \max(\Delta M_b, \phi)$ for each $b$. Similarly to the division constant $\beta$, higher values of $\phi$ imply all vertices (even dissimilar pairs) have higher chances of being chosen. We then randomly draw $b$ according to the created probability distribution.
We present the pseudocode of the re-implemented move function, where we assume $\pi$ is the current permutation represented by the matrix $P$:

\begin{algorithm}
\caption{Move function in guided annealing.}
\begin{algorithmic}[1]
\State $a \gets$ random vertex
\For{every vertex $b$}
    \State $\Delta E_b = max(m_{a,\pi(b)} + m_{b,\pi(a)} - m_{a,\pi(a)} - m_{b,\pi(b)}, \phi)$
\EndFor
\State $\Pi \gets$ probability distribution obtained by normalizing $\Delta E$
\State $b \gets$ vertex randomly drawn from $\Pi$
\State $P' \gets$ permutation generated from $P$ by transposing images of $a$, $b$
\State return $P'$
\end{algorithmic}
\end{algorithm}

\subsubsection{Time complexity of the new version}

The similarity matrix is computed only once at the beginning of a run for each graph, and most of the complexity in this process lies in computing the centralities themselves.

\subsubsection{Optimization of parameters by grid search}

The values of the two introduced parameters, $\beta$ and $\phi$, i.e., the division and probability constants, were determined via grid search on about instances 400 of various random network models with varying sizes (50 to 150 vertices) and parameters. We then selected the parameters with the best average results. When working with PageRank, we also conducted a grid search to optimize the damping factor $\alpha$, but without significant differences in results; therefore, we keep using the default value of $\alpha = 0.85$.
Generally, the few top combinations of parameters performed similarly, whereas the other combinations performed significantly worse. We do not claim the global optimality of the parameters we selected and recognize that different datasets could produce different parameters. However, the parameters we selected were sufficient to answer our hypothesis about guided annealing's improvement in symmetry calculation.

\subsection{Evaluation methodology}

We now describe statistical methods used to evaluate the performance of different annealing versions. Violin plots are utilized to visually represent the distribution of measured symmetry (i.e. values of $S$, where lower values imply higher symmetry) across different instances of the models. Paired $t$-tests are used to assess whether the mean values of symmetry values obtained from two different annealing versions differ. The null hypothesis assumes that there is no significant difference, and the standard significance level of $0.05$ is used to reject the null hypothesis. Finally, Cohen's $d$ quantifies the effect size.

\section{Results}

\subsubsection{Grid}

\begin{figure}[hptb]
    \centering
    \makebox[\textwidth]{\includegraphics[height=15cm, keepaspectratio]{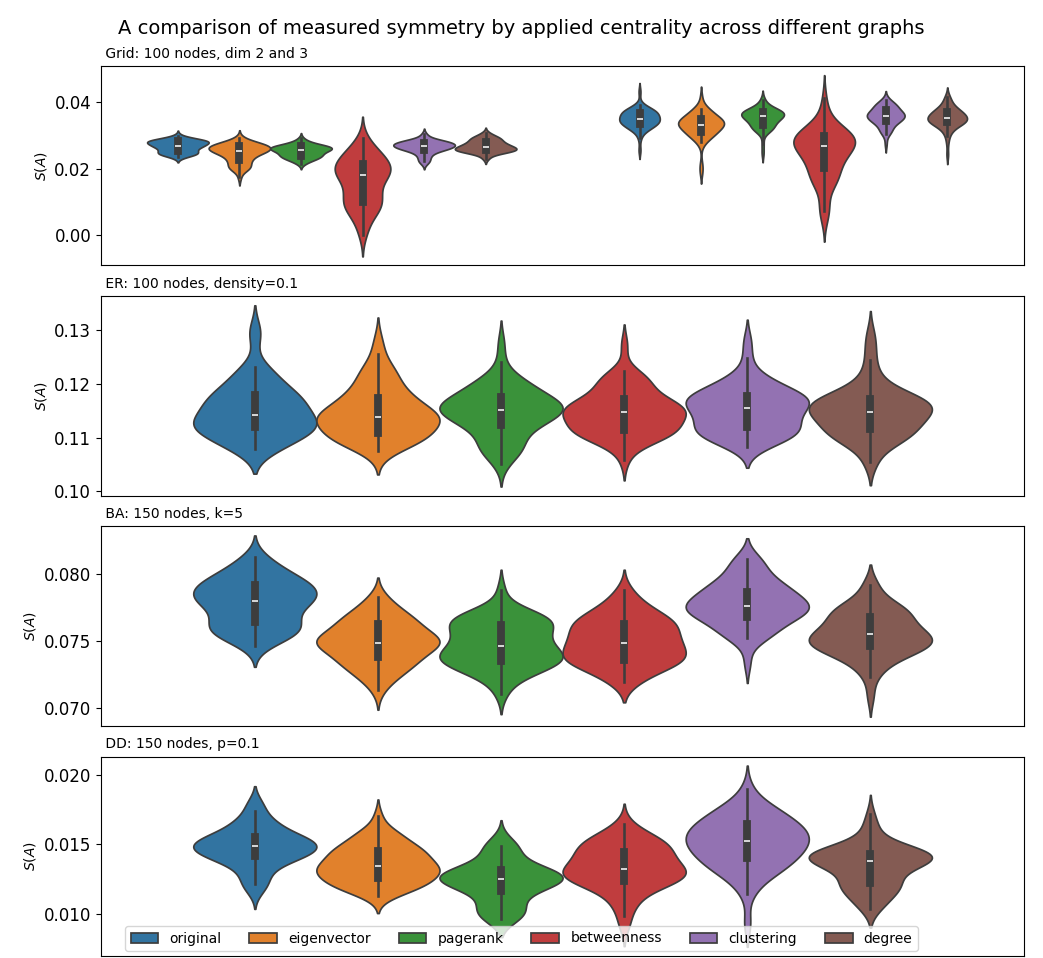}}
    \caption{Comparison of the performance of simulated annealing guided by different centralities across various random network models. The results are visualized in violin plots, where each color represents a version of annealing guided by a different centrality. For each configuration of parameters, we conduct 50 simulations. The first subplot presents results for grid graphs with 100 nodes in 2D and 3D. The second subplot presents results for Erdős–Rényi graphs with 100 nodes and edge density of 0.1. The third subplot presents results for Barabási–Albert graphs with 150 nodes and $k=5$. The final subplot presents results for the Duplication–Divergence model with 150 nodes and a divergence probability of 0.1.}

    \label{fig:Symmetries}
\end{figure}

We evaluate improved annealing on grid graphs with $50, 100, 150$ vertices in 2 and 3 dimensions. In 2D, the lengths of sides are always set to $5 \times x$ and in 3D to $2 \times 5 \times x$. For all grid parameters, guided versions significantly improved computed symmetry, with betweenness yielding the highest improvements, followed by eigenvector centrality (see Fig.~\ref{fig:Symmetries}, the first subplot). 

\subsubsection{Erdős–Rényi model}

We follow with ER graphs with $n = 20, 50, 100$ and edge densities $p = 0.1, 0.3, 0.5$. 
We observe that there were no statistically significant differences between results produced by the different annealing versions, as presented in the second subplot of Figure \ref{fig:Symmetries}.
A possible explanation would be that
the ER model is entirely random; it lacks any internal structure stemming from a generation process and has no predictable substructures that can be reflected in each other. 

\subsubsection{Barabási–Albert model}

We conduct measurements on instances with 50, 100, and 150 nodes and $k$ values of 3, 5, and 7.
The results for BA graphs on 150 nodes and $k = 5$ are presented in the third subplot of Figure \ref{fig:Symmetries}. 
While not significant for $50$ nodes, as graph size increases to $n = 100$ and $150$, the improvements grow to become more significant. Versions of annealing guided by eigenvector centrality, betweenness, and PageRank yield some level of improvement over original annealing.

\begin{figure}[hptb]
\centering
\begin{subfigure}{0.5\textwidth}
    \centering
    \includegraphics[width=\linewidth]{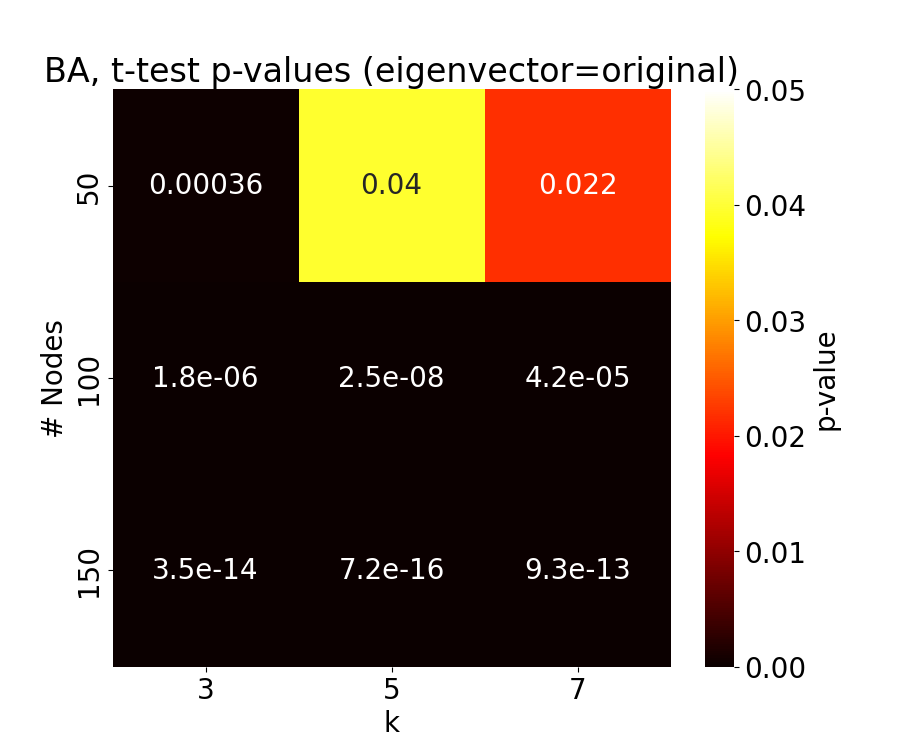}
\end{subfigure}%
\begin{subfigure}{0.5\textwidth}
    \centering
    \includegraphics[width=\linewidth]{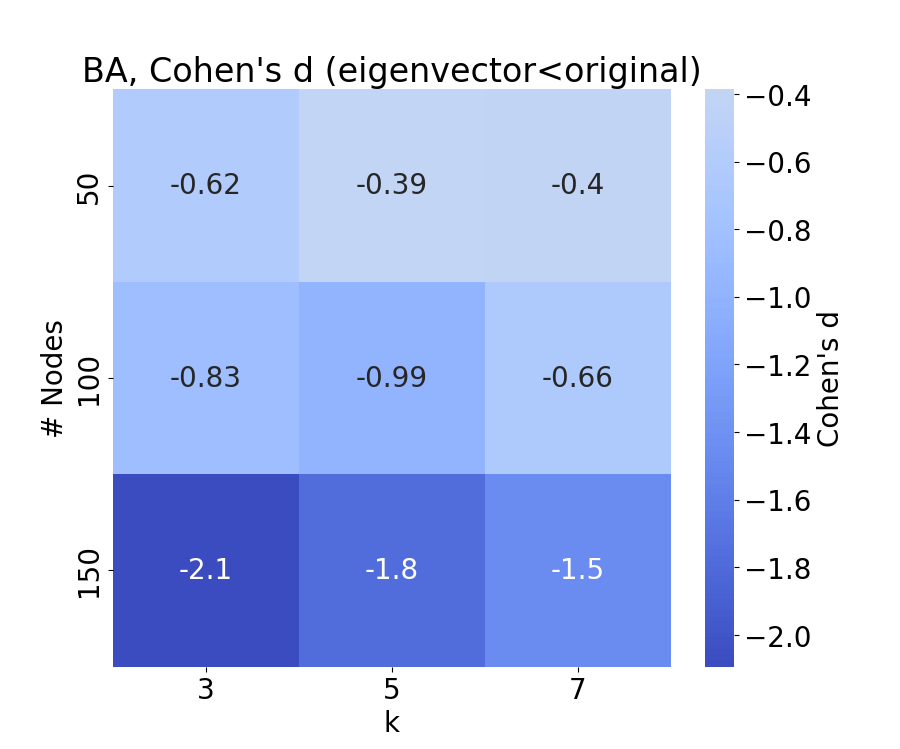}
\end{subfigure}
\caption{Comparison of the performance of original simulated annealing and simulated annealing guided by eigenvector centrality, BA model, where $k$ is the number of links of a new node.}
\label{fig:centralities-BA-p-values}
\end{figure}

We select eigenvector centrality, which by eye seems to perform similarly or better than other centralities, and evaluate its improvements in more detail. We compute paired $t$-tests to determine whether the two algorithms (eigenvector-centrality-guided and original annealing) compute symmetry values with the same mean, along with Cohen's $d$ to measure the effect size (see~Figure \ref{fig:centralities-BA-p-values}).
Apart from one combination of parameters, annealing guided by eigenvector centrality produces significantly better outcomes compared to original annealing. 
We also see that the larger and sparser the graph is, the greater the improvement, as suggested by the increasing absolute value of Cohen's $d$. 

\subsubsection{Duplication–divergence model}

Finally, we conduct measurements on the DD model. The dataset consists of DD graphs with node counts $n = 50, 100, 150$, and divergence probabilities $0.1$ and $0.3$. The results for DD graphs on 150 nodes and divergence probability $0.1$ are presented in the fourth subplot of Figure~\ref{fig:Symmetries}.

Our findings indicate that guided annealing significantly outperforms original annealing. It also holds that instances with divergence probability $0.3$ are less symmetric than those with divergence probability $0.1$, which we attribute to the fact that as $\sigma$ decreases, the graph becomes sparser and starts to resemble trees, which are intuitively somewhat regular (at least in the sense of having many leaf nodes likely sharing similar properties).

\begin{figure}[hptb]
\centering
\begin{subfigure}{0.5\textwidth}
    \centering
    \includegraphics[width=\linewidth]{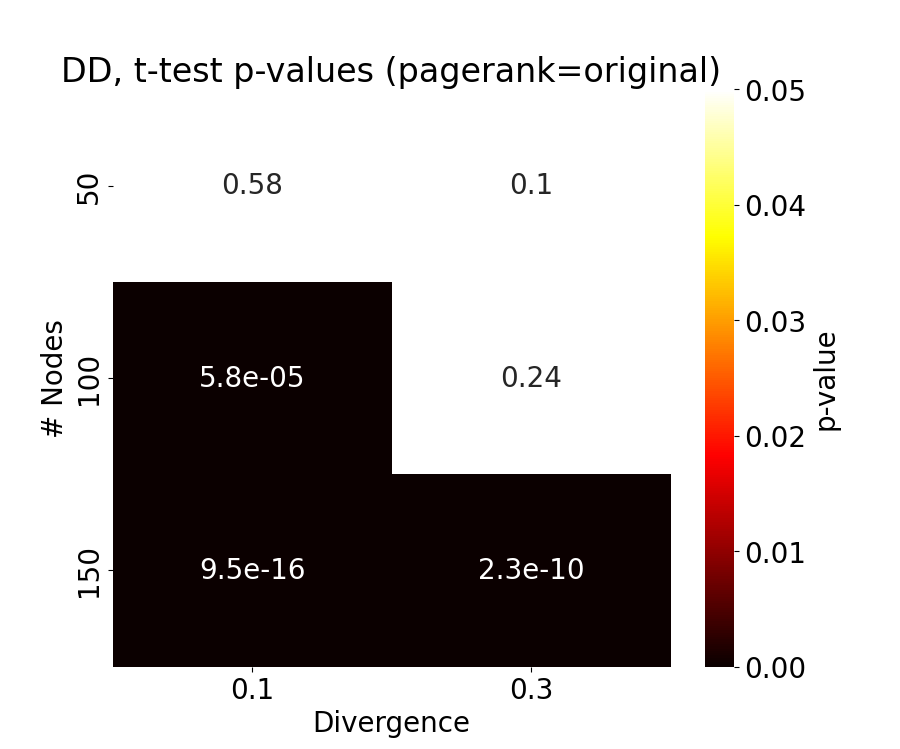}
\end{subfigure}%
\begin{subfigure}{0.5\textwidth}
    \centering
    \includegraphics[width=\linewidth]{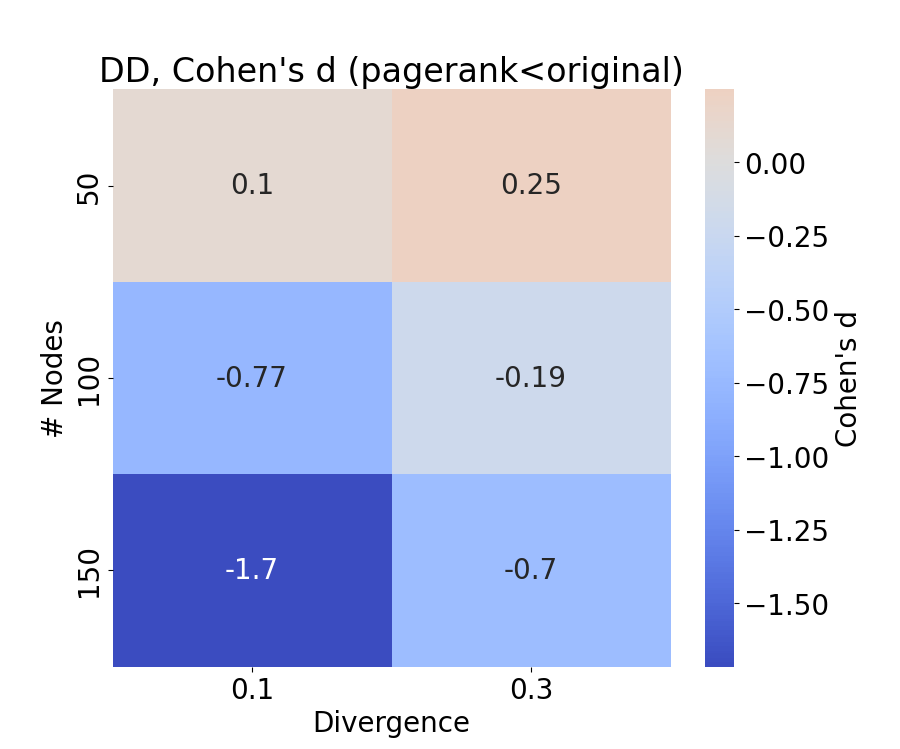}
\end{subfigure}
\caption{Comparison of the performance of original simulated annealing and simulated annealing guided by eigenvector centrality, DD model.}
\label{fig:centralities-DD-p-values}
\end{figure}

 For a more rigorous evaluation, we select PageRank, which seems to perform at least as well as other centralities. Calculating paired $t$-tests and Cohen's $d$ reveals that for smaller graph sizes, we cannot reject the null hypothesis. If we move to larger graph sizes, however, we see PageRank-guided annealing outperforms the original, unguided version, as evident in Figure \ref{fig:centralities-DD-p-values}. The improvements are more significant in sparser DD instances with lower $\sigma$ and also become more pronounced with growing graph size.

\FloatBarrier

\subsubsection{Larger graphs}

\begin{figure}[hptb]
    \centering

    \begin{subfigure}{0.6\textwidth}
        \centering
        \includegraphics[width=\textwidth]{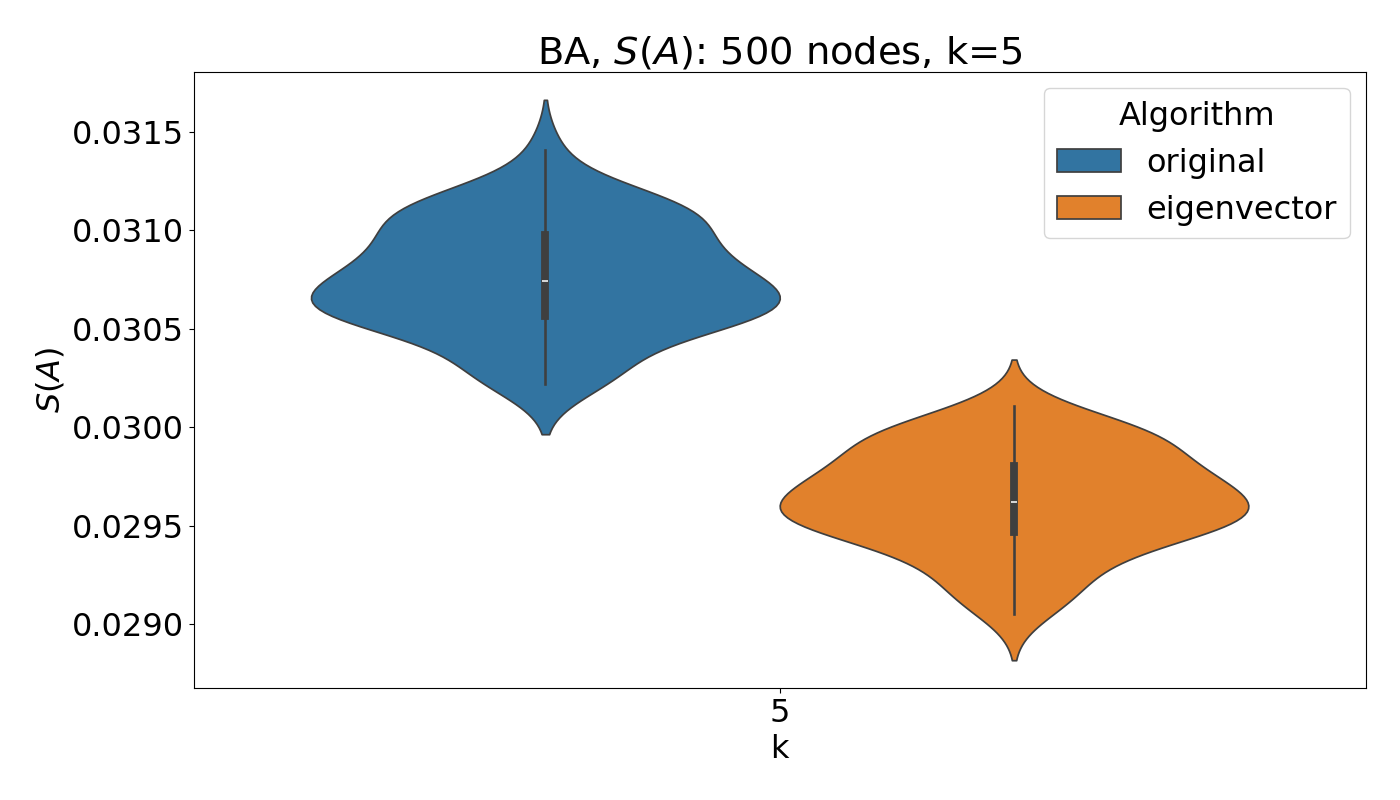}
        \caption{BA graphs: Measured symmetry.}
    \end{subfigure}
    \begin{subfigure}{0.39\textwidth}
        \centering
        \includegraphics[width=\textwidth]{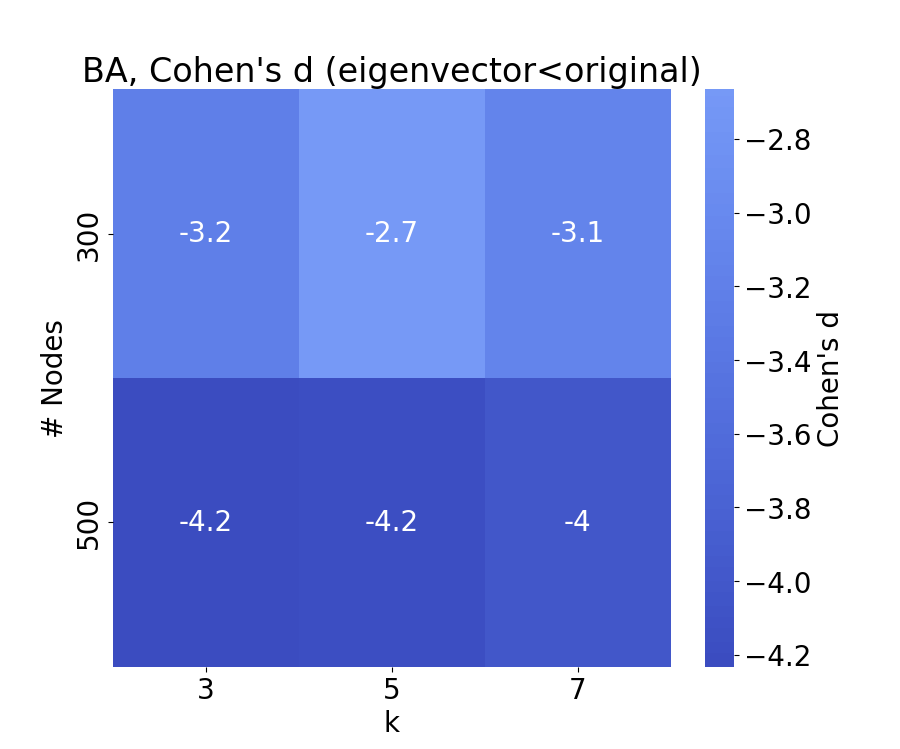}
        \caption{BA graphs: Cohen’s d.}
    \end{subfigure}

    \begin{subfigure}{0.60\textwidth}
        \centering
        \includegraphics[width=\textwidth]{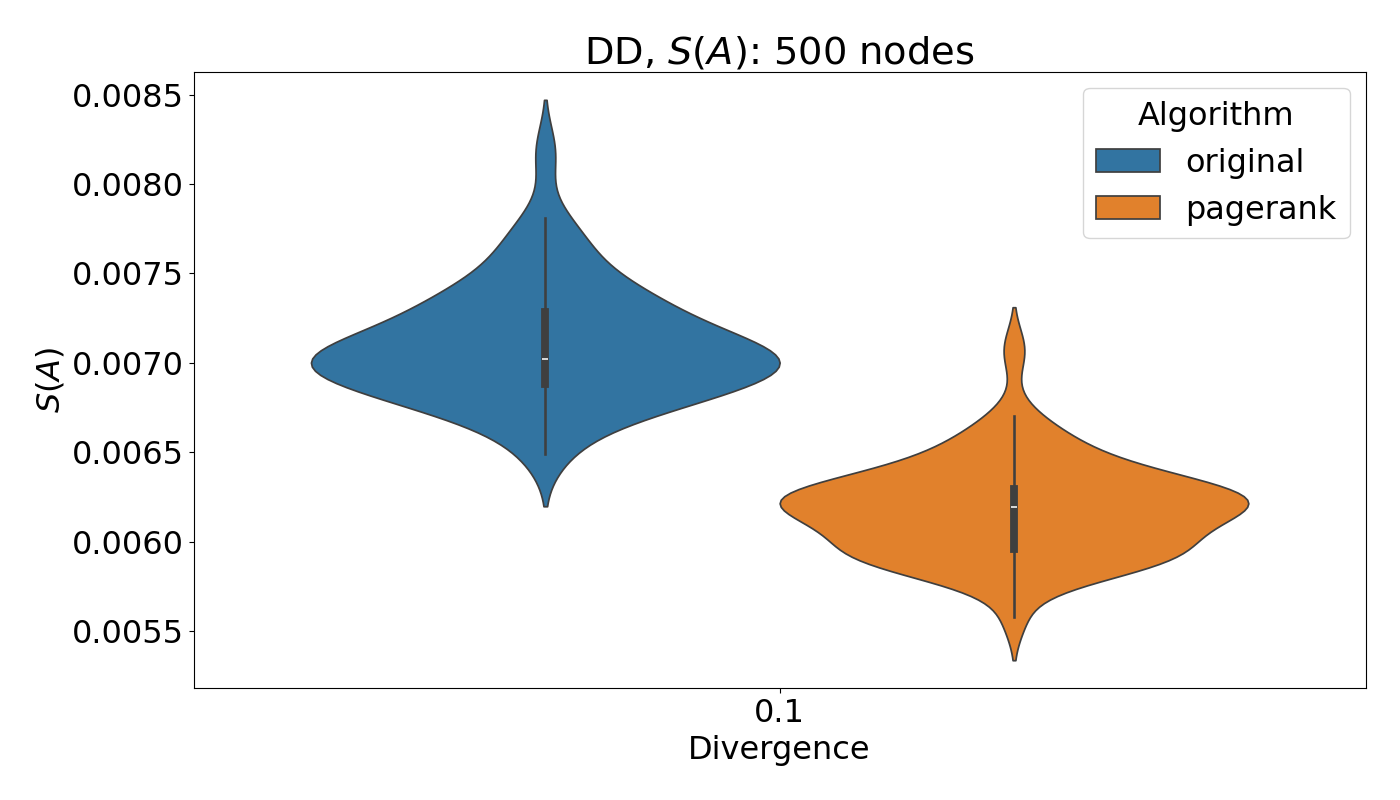}
        \caption{DD graphs: Measured symmetry.}
    \end{subfigure}
    \begin{subfigure}{0.39\textwidth}
        \centering
        \includegraphics[width=\textwidth]{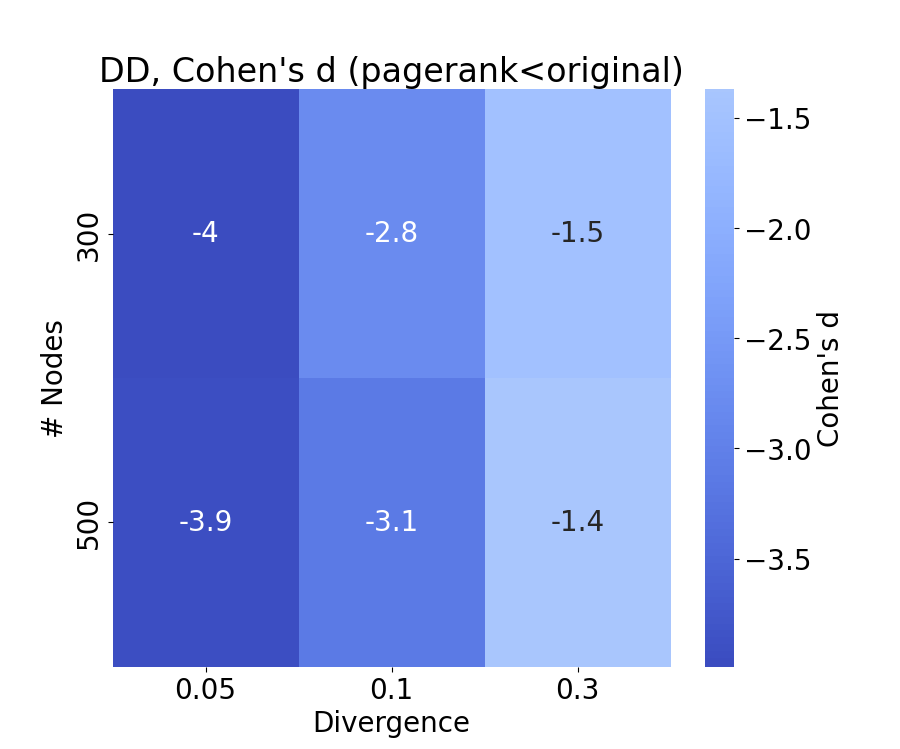}
        \caption{DD graphs: Cohen’s d.}
    \end{subfigure}

    \caption{Comparison of measured symmetry and Cohen’s d values for larger BA and DD graphs.}
    \label{fig:large_graphs_comparison}
\end{figure}

In the previous section, we observed a trend in both the BA and DD models; improvements achieved by guided annealing became more pronounced with growing graph size. To further explore this trend, we extend our analysis to graphs of sizes $300$ and $500$. Specifically, we focus on the performance of PageRank-guided annealing on larger DD graphs, and eigenvector-centrality-guided annealing on larger BA graphs. Figure \ref{fig:large_graphs_comparison} shows the distribution of measured symmetry for $n = 500$ nodes and Cohen's $d$ effect size for other combinations of the parameters. In both cases, the guided versions of the algorithm yield even more significant improvements over original annealing than in smaller graph instances.

\section{Conclusion}

This study investigated the impact of guiding simulated annealing with graph centralities as heuristics to improve the computation of approximate network symmetry. Given that graph centralities are preserved by automorphisms, we hypothesized that aligning vertices with similar centrality values would lead to better values of approximate symmetry.

We introduced a new heuristic algorithm for finding approximate symmetry extending the existing simulated annealing approach. As a heuristic, we used network centralities, namely degree centrality, eigenvector centrality, betweenness centrality, and clustering coefficient. We also defined an efficient way of selecting similar vertices to be transposed. The effectiveness of this new method was evaluated across several graph classes or random models, including the grid, the Erd\H{o}s–R\'{e}nyi random graph, the Barabási–Albert model, and the duplication–divergence model.

Our results demonstrated that centrality-guided annealing improved approximate symmetry calculation concerning the model used. In grid graphs, betweenness centrality significantly improved computed approximate symmetry. This can be explained by the fact that this centrality rigorously considers communication using the shortest paths and in a grid graph this centrality characterizes vertices relatively uniquely. However, this graph should be understood as a test instance of the basic principle. 

In contrast, no significant improvements were observed for Erd\H{o}s–R\'{e}nyi graphs, which we attribute to some of their known properties, e.g. small clustering of ER model implies no improvement using clustering coefficient, whereas degree or even PageRank are potentially confused by many similar nodes. Similarly, given the short distances and generally good connectivity of the ER model, neither betweenness centrality is a suitable discriminating characteristic. This may seem like a disappointing result, however, given the small reproducibility of real networks using the ER model, this result is not crucial.

In the Barabási–Albert and duplication-divergence networks, eigenvector centrality, and PageRank led to measurable improvements, particularly in instances that were inherently more symmetric. With graph size increasing from 100-150 vertices to 300-500 vertices, the improvements in Barabási–Albert and duplication–divergence networks became even more pronounced. These results correspond to intuition in the BA model due to the specific connection of vertices to each other and the creation of a certain hierarchical structure characteristic also typical for real networks~\cite{NetworkScience}. For the duplication-divergence model, the results obtained are more surprising but can be attributed to some structure in the repeated generation of the duplicated vertex. Considering the model parameters, the goal is to test possible further variations in the future. It should also be mentioned that for both networks betweenness centrality works almost similarly to the above-mentioned spectral characteristics. This is probably due to the better resolution of this centrality in terms of its more comprehensive use of more distant areas from the corresponding node. On the other hand, the clustering coefficient performed poorly, likely due to the high locality of this characteristic and the fact that the networks were highly clustered.

There are several ways to extend the results mentioned here. One weakness seems to be the use of simulated annealing itself. Therefore, it would be benefitial to explore other metaheuristic approaches in the context of approximate symmetry computation, such as evolutionary algorithms, and examine whether centrality-based guidance can improve their performance. 

In the BA model, for example, simple degree centrality performs comparably to the more complex eigenvector centrality. Since this centrality is a generalization of the vertex degree, it would be interesting to distinguish for each model the degree of locality needed for the centrality used. Similarly, betweenness centrality could be considered in its more local versions, such as $k$-betweenness~\cite{Borgatti2006graph}.

Instances of graphs that have uniform centralities across vertices, such as regular graphs having identical vertex degrees, can cause a certain problem for the whole algorithm. These graphs can often be very special. However, for possible use in algorithms, it would be beneficial to identify a well-described class of such uniform graphs, especially for the characteristics yielding good results as a heuristic for approximate symmetry computation. In the case of graphs with uniform betweenness centralities, this is ongoing research~\cite{Hartman2024ontheconnectivity,ghanbari2023structure}. The contribution of such a potential classification can be seen in the fact that such betweenness uniform graphs need not be regular.

\subsubsection*{Acknowledgments} This work was supported by the Czech Science Foundation Grant No. 23-07074S.

%
%
%

\bibliographystyle{plainurl}  
\bibliography{references}   

\end{document}